\documentclass[a4paper,11pt]{JHEP3}
\usepackage{amsmath}
\usepackage{epsf, cite, amssymb}
\usepackage{epsfig}
\title{Intersecting D7-Branes, I5-Branes and Conifolds}
\author{Ling-Yan Hung \\ Department of Applied Mathematics and
  Theoretical Physics, \\ Wilberforce Road, Cambridge CB3 0WA, UK
\\ E-mail: \email{lyh20@damtp.cam.ac.uk} }
\abstract{A candidate supergravity solution of intersecting D7-branes with a
five-dimensional intersecting domain (an I5-brane) is presented.
This displays an enhanced Poincare symmetry and supersymmetry away from the
brane cores. We also explore the possibility of a relation between the
intersection region of D7-branes and conifolds through F-theory.}

 \preprint{DAMTP-2006-118}

\begin{document}

\section{Introduction}

The study of different configurations of branes provides us with
information about their dynamics and properties of the gauge theories that reside in them.
For example, by considering space filling
intersecting branes as in \cite{green}, the anomaly inflow mechanism
determines the coupling of D branes to Ramond-Ramond forms.  
As is noted in \cite{green} the two distinct
supersymmetric configurations of space filling intersecting branes up
to T-duality are intersecting D5-branes over $1+1$ dimensions ($I1$ brane) and intersecting
D7-branes over $5+1$ dimensions($I5$ brane). The characteristic
feature of such configuration is that they describe chiral theories in
the intersection domain. There is a curious symmetry enhancement
in the $I1$ observed in \cite{IKS}, where it was found that Poincare symmetry
is enhanced from SO(1,1) to SO(1,2), and the number of supersymmetries was also
doubled. This phenomenon was confirmed both from the weakly coupled
gauge theory picture and the supergravity limit. 
It is thus interesting to see if this also occurs in the $I5$-brane.
However in this case the microscopic description in terms of
six-dimensional gauge theory is problematic. Furthermore
the validity of the use of DBI action in analysing the D7-branes
dynamics is not well understood, since an angle
deficit is created by every 7-brane independent of the value of the
string coupling. 
On the other hand it should be possible to study the supergravity
limit of this configuration. This is one of the main purposes of this paper.

Our solutions generally describe a stack of $N_1$ coincident D7-branes
intersecting another stack of $N_2$ coincident D7-branes in an $I5$
configuration, where $N_1, N_2 \le 24$. Most of our discussion will be
restricted to the case with equal numbers of branes, $N_1=N_2$. The
solution is based on a generalisation of the stringy cosmic string\cite{cosmicstring}.
Interestingly, the solutions also have enhanced Poincare symmetry and
supersymmetry, as in the case of I1 branes. However, in this case the
enhancement is a property of the full solution rather than the
near-horizon limit. The solutions also generalise to situations in
which branes are separated in one of the stacks and also to branes
intersecting at angles.

The second part of this paper explores possible connections between intersecting D7-branes
and the conifold singularity of elliptically fibred Calabi-Yau
manifolds. The fact that a particular system of branes should be dual to
special spacetime geometry has been discussed at length in the
literature (e.g.\cite{Vafa3}). The explicit example of intersecting
NS5-branes which T-dualise to the  conifold provides an
explanation for gauge symmetry enhancement as the complex
moduli of the geometry is tuned. The similarities between these NS5
configurations and intersecting D7-branes suggests an extension of
these ideas might be possible. In particular, F-theory describes
elliptically fibred Calabi-Yau manifolds in terms of two intersecting
stack of D7-branes with 24 branes in each stack. We will show
explicitly that the conifold arises in the limit in which two
D7-branes, one from each stack, intersect each other but are far from
the others.
However, we have not been able to obtain the conifold metric
from our solution, mainly because of the fact that F-theory 
distinguishes the extra fibred torus, which is unphysical, from the rest
of the spacetime, thus breaking the SO(4) symmetry required of a
conifold. 

The organisation of the paper is as follows: 
The solution for parallel D7-
branes and its origin from the stringy cosmic string solution 
will provide the framework for the intersecting brane solution. 
The stringy cosmic string\cite{cosmicstring} and its connection to
D7-branes and F-theory, which foreshadows the analysis of the
conifold, are reviewed in the appendix.
In Section 2 and 3 we shall present the candidate solutions of the intersecting
D7-branes and some of their properties. In section 4  we will discuss
the possibility of obtaining a conifold at the
intersection. Further comments will be made in section 5.

\section{Intersecting D7-brane solutions of the Einstein Equations}
The bosonic action and the equations of motion are as discussed in
appendix (\ref{cosmicsolun})
To obtain a solution that represents a system of two sets of D7-branes intersecting in six spacetime dimensions with four relatively
transverse directions, we assume that the metric takes the following form:

\begin{equation}
ds^2 = \eta_{\alpha\beta}dx^\alpha dx^\beta + K_{a\bar{b}}dz^a d\bar{z}^b
\end{equation}
where $\alpha, \beta \in \{0,1,2,3,4,5\}$ and $a, b \in \{1,2\}$ such that $z^a$'s are two complex dimensions.
In other words, the four relatively transverse directions form a complex two dimensional Kahler manifold whose Kahler metric
is given by $K_{a\bar{b}} = \partial_a\bar{\partial_b}K$, where $K$ is the Kahler Potential.

The $\tau$ equation (\ref{tau}) can be solved if $\tau$ is a holomorphic function of $z^a$. Then the Einstein equation (\ref{einstein})
is simplified to
\begin{equation}\label{simpleeinstein}
R_{a\bar{b}} =  -\partial_a \partial_{\bar{b}} \ln \textrm{det}K_{a\bar{b}} = \frac{1}{4\tau^2_2}(\partial_a\tau \partial_{\bar{b}}\bar{\tau}) = -\partial_a \partial_{\bar{b}} \ln \tau_2
 \end{equation}
where we have made use of the property of a Kahler manifold in the first equality.

The equation is solved by
\begin{equation}\label{solution1}
\tau= \tau(z^a),
\end{equation}
\begin{equation}\label{solutiondet}
\textrm{det} K_{a\bar{b}} = |F(z^a)|^2 \tau_2
\end{equation}
where $F(z^a)$ is any arbitrary holomorphic function.

SUSY transformations of the dilatino $\lambda$ and gravitinos $\psi_\mu$ are given by:
\begin{equation}
\delta \lambda = -\frac{1}{2\tau_2}(\partial_a \tau e^a_A \Gamma^A + \partial_{\bar{a}}\tau e^a_A \Gamma^{\bar{A}} ) \epsilon^*
\end{equation}
\begin{equation}
\delta\psi_a = \textrm{D}_a \epsilon = (\partial_a + \frac{1}{4}w^{AB}_a\Gamma_{[AB]} + \frac{i}{4} \frac{\partial_a\tau_1}{\tau_2}) \epsilon 
\end{equation}
where $w^{AB}_a$ are the spin connections (whose non-trivial components are along the relatively transverse directions only) and $A,B \in \{1,2\}$ are tangent space indexes.
Both variations vanish when we substitute in the solution
(\ref{solution1}), (\ref{solutiondet}) and only one quarter of maximal
supersymmetries are preserved.

\subsection{An ansatz for coincident D7-branes intersecting orthogonally}\label{basic}

To obtain an exact form of the metric, we make further assumptions of the form of the holomorphic function $F$ and $\tau_2$:
suppose we are having $N_1$ coincident D7-branes at the origin transverse to directions $z$ intersecting another set of $N_2$ D7-branes also at the origin transverse to directions $w$, then we propose that   
\begin{equation}\label{jeq}
j(\tau) = \frac{1}{z^{N_1}w^{N_2}} ,
\end{equation}  
\begin{equation}
F(z,w) = \eta^2(\tau) (z^{ -\frac{N_1}{12}}w^{-\frac{N_2}{12}}).
\end{equation}
This proposal is a direct generalisation of the solution for parallel D7-branes. 
The solution restricts $\tau$ in the fundamental region and ensures modular invariance of the metric. The factor $z^{ -\frac{N_1}{12}}w^{-\frac{N_2}{12}}$ cancels the divergence in $\eta^2(\tau)$ as $z$ and $w$ approach $0$.

In this proposal, both $F$ and $\tau$ depend only on $z^{N_1}w^{N_2}$ and this suggests that we should define new variables:
\begin{eqnarray}\label{xy}
y= -(N_1\ln z + N_2 \ln w) &,& x= (N_2\ln z - N_1 \ln w).
\end{eqnarray}
In these new variables, equation (\ref{solutiondet}) becomes 
\begin{equation}
\textrm{det}K_{z\bar{w}} = \left| \begin{array}{cc}
-\frac{N_1}{z} &  -\frac{N_2}{w} \\
\frac{N_2}{z} &  -\frac{N_1}{w}
\end{array} \right| ^2 (K_{x\bar{x}}K_{y\bar{y}} - K_{y\bar{x}}K_{x\bar{y}}) = |F|^2 \tau_2.
\end{equation}

Rearranging gives 
\begin{equation}\label{eom}
(K_{x\bar{x}}K_{y\bar{y}} - K_{y\bar{x}}K_{x\bar{y}})  = \frac{\exp\left(\frac{N_2-N_1}{N_1^2+N_2^2}(x+\bar{x})\right)\exp\left(-\frac{N_2+N_1}{N_1^2+N_2^2}(y+\bar{y})\right)}{(N_1^2+N_2^2)^2}|F|^2 \tau_2,
\end{equation}
using the fact that
\begin{equation}
|zw|^2 = \exp\left(\frac{N_2-N_1}{N_1^2+N_2^2}(x+\bar{x})\right)\exp\left(-\frac{N_2+N_1}{N_1^2+N_2^2}(y+\bar{y})\right).
\end{equation}
The simplest solution would be to require that the metric be block diagonal in $x$ and $y$ i.e. $K_{x\bar{y}}=K_{y\bar{x}}= 0$ such that the Kahler Potential takes the form
\begin{equation}
K = f(x,\bar{x}) + g(y,\bar{y}).
\end{equation}
The resultant metric is
\begin{equation}\label{Kxx}
K_{x\bar{x}} = \alpha \exp\left(\frac{N_2-N_1}{N_1^2+N_2^2}(x+\bar{x})\right)
\end{equation}
\begin{equation}\label{Kyy}
K_{y\bar{y}} = \beta \exp\left(-\frac{N_2+N_1}{N_1^2+N_2^2}(y+\bar{y})\right)|F|^2 \tau_2,
\end{equation}
where $\alpha,\beta$ are constants satisfying $\alpha \beta = 1/(N_1^2
+ N_2^2)^2$. 
The choice of $\alpha$ and $\beta$ has the
effect of altering the way the arguments of $z,w$ are mixed with each
other. Consider for example the case where $N_1 = N_2 = 12$
where both $K_{x\bar{x}}$ and $K_{y\bar{y}}$ become flat. The compact
part of the line element in terms of the original coordinate  is
(i.e. for $z = |z|\exp(i\theta)$ and $w = |w|\exp(i\phi)$)
\begin{equation}\label{compact_part}
ds_{\theta,\phi}^2 = \alpha d(\theta - \phi)^2 + \beta d(\theta + \phi)^2.  
\end{equation}
This represents different ways of fibering an $S_1$ over another
$S_1$. The fibration becomes trivial only when
$\alpha = \beta$. 

Consider however the case when $N_2 = 0$ (i.e. one single
stack of parallel branes), our solution 
reduces to the solution of single set of parallel 7-branes by
transforming back to the $z$ and $w$ coordinates, and rescaling $z =
(\tilde{z}/(N_1\sqrt{\beta}) $, and similarly 
$w = (\tilde{w}/(N_1\sqrt{\alpha})$. The solution of parallel
D7-branes is as expected independent of the choice of $\alpha$.

\subsubsection{Different choices of coordinates}
Our ansatz for solving the Einstein equations requires us to define a new set
of coordinates $x,y$ such that all non-trivial coordinate dependence
is in $y$. This suggests that while the choice of $y$ is unique up to
an arbitrary rescaling, we have far more freedom in choosing
$x$. A more general choice of $x$ would be
\begin{equation}
x = A \ln z + B\ln w,
\end{equation}
where $AN_2 -BN_1\ne 0$ so that the Jacobian is non-singular. This change
in the choice of coordinates is inequivalent to a mere
coordinate transformation on the resultant metric and are genuinely
different solutions. We therefore want to  know what sort of geometry
is obtained, even
though we started off 
with the same ansatz for $\tau$. Solving the Einstein equation
as before with this different choice of coordinate, gives
\begin{eqnarray}
K_{x\bar{x}} &=& \alpha
\exp{(\frac{N_2-N_1}{N_2A-N_1B}(x+\bar{x}))} \nonumber \\ 
K_{y\bar{y}} &=& \beta
\exp{(\frac{B-A}{N_2A-N_1B}(y+\bar{y}))}|\eta^2\exp({\frac{y}{12}})|
^2 \tau_2 ,  
\end{eqnarray}
where in this case $\alpha \beta = 1/(AN_2 -BN_1)^2$.
For $N_1=N_2=N$ we see that the dependence on $A$ and
$B$ drops out from the exponential, though this is not true for $N_1\ne
N_2$. This again implies that the solutions for  $N_1\ne N_2$ are
qualitatively different from those when $N_1=N_2$. Another issue however is
that even for $N_1=N_2$, this different choice of coordinate still has
a non-trivial effect on the boundary conditions on the angles. This is
most clearly exhibited in the case where $N_1 = N_2 = 12$. In that
case both $K_{x\bar{x}}$ and $K_{y\bar{y}}$ become flat and naively we
would get two cylinders. However similar considerations as in
eq.(\ref{compact_part}) in the previous section implies that the
compact part of the line element has non-trivial dependence on $A$ and
$B$ i.e.
\begin{equation}
ds_{\theta,\phi} \sim \beta d(\theta+\phi)^2 + \alpha d(A\theta + B\phi)^2.
\end{equation}
In effect we again have two entangled cylinders.

\subsection{More general solutions}

\subsubsection{Separated branes}
We would like to generalise the solutions we have obtained so far to
the case where instead of coincident branes we have branes separated
in each stack. The obvious ansatz for $\tau$ would be
\begin{equation}\label{sep}
j(\tau) = \prod_{i,j}\frac{1}{(z-a_i)(w-b_j)}.
\end{equation}
However, with this ansatz the coordinate transformation we have been
using would produce a Jacobian that is no longer expressible as a
product of $f(x)g(y)$ for some holomorphic functions $f,g$ and so
we could no longer obtain block diagonal metric in $x$ and $y$. There
is no obvious way to get around with this problem.
On the other hand, it is possible to solve the Einstein equations when
only one stack of branes separates. In that case 

\begin{equation}
j(\tau)= \prod_{i}^{N_2}\frac{1}{z^{N_1}(w-b_i)}
\end{equation} 
and the choice
\begin{eqnarray}
y &=& -[N_1\ln z + N_2 \ln w] \nonumber \\
x &=& B\ln w.
\end{eqnarray}
gives
\begin{eqnarray}
K_{x\bar{x}} &=& \alpha \exp(\frac{x+\bar{x}}{B}) \prod_i^{N_2}|\exp{(\frac{x}{B}-b_i)}|^{-\frac{2}{N_1}} \nonumber \\
K_{y\bar{y}} &=& \beta \exp(-\frac{y+\bar{y}}{N_1})|\eta^2\exp(\frac{y}{12})|^2\tau_2,
\end{eqnarray}
where $\alpha \beta = 1/(N_1B)^2$ here.
At distances much greater than the separation of the
branes, the solution reduces to that of coincident branes, as should
be expected. 

\subsubsection{Branes intersecting at angles}
To describe N branes intersecting at angles, the ansatz for $\tau$ would
be
\begin{equation}
j(\tau) = \prod_i^N\frac{1}{z-a_iw}.
\end{equation}
The relevant choice of $(x,y)$ coordinates is now
\begin{eqnarray}
y &=& -[N\ln w + \sum_i (\ln (\frac{z}{w}-a_i))] \nonumber \\
x &=& B\ln \frac{z}{w}.
\end{eqnarray}
Fortunately the Jacobian of this coordinate change is $|BN/zw|^2$ and
we can easily repeat the same procedure for solving the Einstein equations to
get
\begin{eqnarray}
K_{x\bar{x}} &=& \alpha \exp(\frac{x+\bar{x}}{B}) \prod_i^{N}|\exp{(\frac{x}{B}-b_i)}|^{-\frac{4}{N}} \nonumber \\
K_{y\bar{y}} &=& \beta
\exp(-2\frac{y+\bar{y}}{N})|\eta^2\exp(\frac{y}{12})|^2\tau_2, 
\end{eqnarray}
and $\alpha \beta = 1/(BN)^2$ here.
The solution is locally identical to the above solution for separated branes
in one of the two stacks with $N_1/2 = N$, except they satisfy different boundary conditions, meaning that the compact
direction corresponding to the angles are entangled in a different
manner in the two solutions. 

\subsubsection{Solutions not block-diagonal in $x,y$}\label{generalsols}
So far we have only solved the Einstein equations assuming that the metric is block diagonal in $x,y$ so that we can easily construct the corresponding Kahler potential. However, there is actually another class of solutions whose Kahler potentials also take a simple form. 
Consider a Kahler potential of the form
\begin{equation}
K = g(y,\bar{y}) + h(x,\bar{x})k(y,\bar{y}).
\end{equation}
The determinant of the resultant metric is
\begin{equation}
(g_{,y\bar{y}}+h k_{,y\bar{y}})h_{,x\bar{x}}k - |h_{,x}k_{,\bar{y}}|^2.
\end{equation}
If we could arrange $K$ such that $hk k_{,y\bar{y}}h_{,x\bar{x}} - |h_{,x}k_{,\bar{y}}|^2 = 0$ then the determinant would become $g_{,y\bar{y}}h_{,x\bar{x}}k$ which is again a simple product of a function in $y$ and $x$ respectively and should solve the Einstein equations. There are two simple distinct cases where this could be achieved. We could have
\begin{equation}\label{new1}
h(x,\bar{x})k(y,\bar{y}) = |m(x)n(y)|^2,
\end{equation}
for any holomorphic function $m,n$. 

A second choice would be to put
\begin{eqnarray}
h(x,\bar{x}) &=& [m(x) \pm m(\bar{x})]^A, \nonumber \\
k(y,\bar{y}) &=& [n(y) \pm n(\bar{y})]^B,
\end{eqnarray}
such that $A+B = 1$.

To prevent extra factors of $x$ appearing in the determinant
of the metric other than the terms already appearing in the Einstein
equation (\ref{eom}), $h$ has to be quadratic in $x$ in the case $N_1=N_2$.  For $N_1\ne N_2$ the component $K_{x\bar{x}}$ has an exponential factor and that seems to dictate that we should choose the first set of solutions ($\ref{new1}$). As we shall see later this new set of solutions is perhaps necessary in helping us to obtain a metric of the conifold close to the intersection region of two orthogonal D7-branes.

\section{A few observations about the solution}
We would now like to explore further some of the implications of our
solution.  We will concentrate on the case where
$N_1 = N_2$ with the standard choice of $x = \ln z - \ln w$ and $y =
-(\ln z + \ln w)$, which was the basic choice made in section (\ref{basic}).

\subsection{Angle Deficit}\label{angledeficit}
Superficially the solution in section (\ref{basic}) looks like a
single D7-brane in the $(x,y)$ coordinate system. However, the
transformation (\ref{xy}) implies that the $x,y$ coordinates
have non trivial boundary conditions and do not span the complex
planes, since $z,w$ do. 
Using the definition (\ref{xy}) and
considering the simple case where $N_1=N_2=N$, we find 

\begin{equation}
y_2 = -(\theta + \phi) , \qquad x_2 = (\theta - \phi),
\end{equation}
where we write 
\begin{equation}
y=y_1+iy_2 , \qquad x=x_1+ix_2 \nonumber 
\end{equation}
and
\begin{eqnarray}
z=|z|e^{i\theta} &,& \qquad w=|w|e^{i\phi},
\end{eqnarray}
and for simplicity we have scaled away the factor of $N$.

Since $\theta$ and $\phi$ are periodic with period $2\pi$, the
fundamental domain in the $(x_2,y_2)$ plane can be chosen to be a
parallelogram with base $4\pi$ and height $2\pi$.  
Since the periodicities of $x_2$ and $y_2$ are linked 
we have to specify the path we are traversing before we can
discuss the angle deficit. Furthermore, there is no meaning to the
number of times the $y$-plane wraps around the fundamental region in
the calculation of the energy of the system.
We also see that for
general values of $N_1$ and $N_2$ the size and shape of the fundamental
region are different from the case where $N_1=N_2$, and this suggests
that the solutions for the former case are qualitatively different
from the latter.
In order to account for the correct periodicities we shall use the coordinates $\theta,\phi$ instead of $x_2,y_2$. Although this complicates the expression for $\tau$, the metric nevertheless simplifies in the near horizon limit (i.e. when $z,w \to 0$) This will be applied in the
calculation of the energy of the system in a later subsection. 

Returning to the discussion of angle deficit, the geometry of an
individual D7-brane is associated with a deficit angle of $\pi/6$
asymptotically far from its core. However, as described in
\cite{cosmicstring} for stringy cosmic string, the solution is smooth
close to the core. The geometry with the intersecting D7-branes is more subtle since each D7-brane affects the world-volume
of the other. We would expect the angle deficit
given by our solution of
intersecting D7-branes to reduce to that of parallel D7-branes by taking first
the limit to a region far away from one of the stacks
of branes and then moving asymptotically far away from the second stack
of branes. The angle deficit is computed by moving around the second set
of branes in the plane transverse to it (say the $z$-plane) while keeping
$w$ fixed. In such a case we would be traversing simultaneously the $x$ and
$y$ plane. 
In this limit 
\begin{equation}
K_{y\bar{y}} = \beta \exp\left(-\frac{N_2+N_1}{N_1^2+N_2^2}(y+\bar{y})\right)|F|^2 \tau_2 \sim \exp\left((\frac{1}{12}-\frac{N_2+N_1}{N_1^2+N_2^2})(y+\bar{y})\right).\end{equation}
By a further coordinate transformation given by
\begin{eqnarray}\label{sigma}
\alpha \exp\left(\frac{N_2-N_1}{N_1^2+N_2^2}(x+\bar{x})\right) dxd\bar{x} &=&
\alpha \exp\left(P(x+\bar{x})\right) dxd\bar{x} = \alpha \frac{d\sigma
  d\bar{\sigma}}{P^2} \nonumber \\
\beta
\exp\left((\frac{1}{12}-\frac{N_2+N_1}{N_1^2+N_2^2})(y+\bar{y})\right)dyd\bar{y}&=&
\beta \exp\left(Q(y+\bar{y})\right)dyd\bar{y} = \beta \frac{d\eta d\bar{\eta}}{Q^2}
\end{eqnarray} 
the angle deficits in the $\eta$ plane is exactly what one would expect
for parallel coincident D7-branes. For $N_1=N_2$ the $x$ plane becomes
a cylinder without the need for a further coordinate
transformation. In general the  effect of the orthogonal set of
branes cannot be removed by moving away from them.

Now consider a path that is simultaneously far
from both stacks of branes and encircles the intersection region. Such a path keeps $x$ fixed and traverses a circle in the $y$-plane. By considering again
the simple case $N_1=N_2=N$, the
path traverses through an angle
of $2\pi$ in both the $z$-plane and the $w$-plane, resulting in an
angle deficit in the $\eta$ plane of $2N\pi/6$, which is the
combined effect of $2N$ branes. 
In particular, it should be noted that when $N_1=N_2=12$ both the $x$
and $y$ plane become cylinders and each plane has an angle deficit of
$2\pi$, independent of the path we take. As remarked these cylinders
are non-trivially 
fibred and only disentangle when $\alpha = \beta$.

\subsection{Monodromies}
The discussion in the previous section carries over to the analysis of
monodromies.
Consider a closed path traversing the $z$ plane with $w$ fixed. This
translates into the $x,y$ coordinates as a closed curve about $x$ and
$y$ simultaneously. When we are sufficiently close to the stack of
$N_1$ branes transverse to the $z$-plane then $C_0 =
\tau_1 \sim \frac{N_1}{2\pi}\theta_z$ and we get exactly the monodromy
we would expect around $N_1$ branes. 

Conversely if we consider moving in a closed curve in $y$ keeping $x$ fixed we would be moving in a curve which simultaneously moves around the $z$ and $w$ plane.
If we integrate the RR 0-form field strength about a closed curve in the
$y$ plane in the region where $|y|$ goes to zero, we get
\begin{equation}
\oint \partial_{y_2} C_0 dy_2
\sim \int^{2\pi (N_1+N_2)}_0 \frac{-1}{2\pi} dy_2 = -(N_1+N_2),
\end{equation}
since $C_0 = \tau_1 \sim \frac{-1}{2\pi}y_2$ as $|y|$ approaches zero. This suggests a non-trivial combination of the monodromies when we move around both stacks of branes along a curve corresponding to constant $x$. The resultant monodromy is that of a stack of $N_1+N_2$ D7-branes transverse to $y$.

Note that while $C_0$ depends only on one variable $y$, there are
in fact more than one homotopically distinct classes of loops. This
is because when we try to smoothly deform a set of loops that circles
one set of the branes to loops circling
the other set, we inevitably hit singularities corresponding to the
brane core (i.e. at $z=0$ and $w=0$). This is
analogous to the case with N parallel but
non-coincident 7-branes. In that case everything depends only on one
complex coordinate $z$ but there are singularities at $N$ different
points in the complex plane and there are $N$ homotopically
inequivalent classes of loops.

\subsection{Issues concerning enhanced supersymmetry}
For two D7-branes intersecting over six spacetime dimensions, we
should expect only a quarter of supersymmetry to be preserved. Yet the
metric in the $x$ plane can always be made flat suggesting that the
amount of supersymmetry is doubled everywhere, contradicting the
prediction of string perturbation theory. This is, however, an
illusion. This is related to the non-trivial boundary conditions in
the $(x,y)$ coordinate system as discussed earlier. To make this
explicit, consider the Killing spinor equation\footnote{We are
  adopting the notations in \cite{Brandeis}.}
\begin{equation}
\delta\psi_a = \textrm{D}_a \epsilon = (\partial_a +
\frac{1}{4}w^{A\bar{B}}_a\Gamma_{[A\bar{B}]} +
\frac{1}{4}w^{\bar{A}B}_a\Gamma_{[\bar{A}B]} + \frac{i}{4}
\frac{\partial_a\tau_1}{\tau_2}) \epsilon = 0
\end{equation}
where $w^{A\bar{B}}_a$ are the spin connections (whose non-trivial components are along the relatively transverse directions only).  $A,\bar{B} \in \{1,2\}$ are complex tangent space indexes while $a,\bar{b}$ are complex spacetime indexes corresponding to $z$ and $w$.
We need the projections
\begin{eqnarray}
\Gamma^{\bar{A}}\epsilon = 0 &,& \Gamma^{A}\epsilon^* = 0.
\end{eqnarray}
This ansatz thus explicitly preserves only a quarter
supersymmetry. Yet, the trivial dependence on the $x$ coordinate
implies that its corresponding spin-connection should vanish, 
rendering the projections along $x$ unnecessary. As a
result supersymmetry is seemingly enhanced. We will show that the non-trivial
boundary conditions inherited from the coordinate transformation
ensures that supersymmetry is enhanced only locally, but not
globally. 

Using properties of Kahler manifolds,
\begin{equation}
\sum_A {w_a^A}_A = -\partial_a \ln \det \bar{e}^{\bar{B}}_{\bar{b}}
\end{equation}
where $e^{B}_{a}\bar{e}^{\bar{B}}_{\bar{b}} = K_{a\bar{b}}$,
the killing equation reduces to
\begin{equation}\label{solution}
\partial_a \ln \det \bar{e}^{\bar{B}}_{\bar{b}} = \frac{1}{2}\partial_a \ln \tau_2.
\end{equation}
whose solution is 
\begin{equation}
\textrm{det} K_{z\bar{w}} = |F(z,w)|^2 \tau_2,
\end{equation}
which is automatically a solution to the Einstein equation
(\ref{simpleeinstein}).
After the coordinate transformation specified in (\ref{xy}) this becomes
\begin{equation}\label{340}
\textrm{det} K_{x\bar{y}} = |zw|^2|F(x,y)|^2 \tau_2.
\end{equation}

The factor of $|zw|^2$ originates from the Jacobian of the
transformation. In the presence of this term, eq (\ref{340}) solves eq
(\ref{solution}) everywhere except at the origin. This can be made
explicit by substituting eq(\ref{340}) into (\ref{solution}) which gives
\begin{equation}
\partial_a \ln \det \bar{e}^{\bar{B}}_{\bar{b}} = \frac{1}{2}\partial_a \ln \tau_2 + \partial_a \ln \bar{z}\bar{w}.
\end{equation}
This extra term is zero everywhere except at $z =0$ or  $w=0$. This is
analogous to a conical space whose metric 
$ds^2 = |z^n|^2 dz d\bar{z}$
can be made flat locally by the transformation
$z^ndz = d\eta$. 
The coordinate transformation is singular at the origin and the spin
connection is proportional to $\partial_z \ln \bar{z}$ as in our
solution. As is well known a conical space in general breaks all
supersymmetry. For our solution, it takes us back to a quarter
supersymmetric state. 
In other words, we should evaluate the spin connection in the $z,w$
coordinates, which makes it explicit that supersymmetry is  enhanced
everywhere except 
at $z,w=0$.

\subsection{Energy of the system - the ADM mass}
For a single cosmic string, the integral
\begin{equation}
E = \frac{1}{2}\int d^2z \sqrt{-g} R
\end{equation}
is related to the energy of the system. In the $x,y$ coordinates the
space is asymptotically flat and we can define the ADM mass as
usual. The above relation is obtained by using the $R_{00}$ component
in the Einstein equation which relates it to the energy momentum
tensor. The result of the integral can be found by changing coordinates from $(x_1,x_2,y_1,y_2)$ to $(x_1,y_1,\theta,\phi)$ to avoid the awkward identification of the
imaginary parts of $x$ and $y$. 
Since the case where $N_1 \neq N_2$ is qualitatively different from the case $N_1 = N_2 = N$, we consider here only the latter situation.
\begin{equation}
E = \frac{1}{2}\int \sqrt{-g} R d^2xd^2y = \frac{1}{2} \int \sqrt{-g} \partial_a\bar{\partial_b}[\ln(\det g_{m\bar{n}})] g^{a\bar{b}}  d^2xd^2y.
\end{equation}
Integrating by parts, we have
\begin{equation}\label{energy}
\frac{1}{2}\int d^2xd^2y   \partial_a (\sqrt{-g}\bar{\partial_b}\ln H g^{a\bar{b}}) - \bar{\partial_b}\ln H \partial_a(\sqrt{-g}g^{a\bar{b}}),
\end{equation}
where $H = \det g_{m\bar{n}} = \tau_2|\eta^2(\tau)e^{(\frac{N}{12}(y+\bar{y})-1)}|^2 $. 

The second term vanishes. Since for a tensor of weight one the covariant derivative acts by
\begin{equation}
\nabla_a \tilde{T}^{a\bar{b}} = \partial_a \tilde{T}^{a\bar{b}} + \Gamma^a_{am}\tilde{T}^{m\bar{b}} + \Gamma^{\bar{b}}_{am}\tilde{T}^{am} - \Gamma^m_{ma} \tilde{T}^{a\bar{b}}
\end{equation}
and $ \Gamma^{\bar{b}}_{am}$ is zero for a Kahler metric. 
Using this relation we can write eq.(\ref{energy}) as

{\setlength\arraycolsep{2pt}}
\begin{eqnarray}\label{energy2}
E&=&\frac{1}{4}\int \partial_x[(\sqrt{-g}g^{x\bar{x}})\bar{\partial_x}\ln
  H] + \bar{\partial_x}[(\sqrt{-g}g^{x\bar{x}})\partial_x \ln H] +{}
\nonumber\\
&& +\partial_y[(\sqrt{-g}g^{y\bar{y}})\bar{\partial_y}\ln H] + \bar{\partial_y}[(\sqrt{-g}g^{a\bar{b}})\partial_y \ln H ].
\end{eqnarray}

Then we express the integrand in terms of the real parts of $x$ and $y$ i.e. $x_1$ and $y_1$. For the imaginary parts we revert to the original angular variables of $z$ and $w$. i.e.
\begin{equation}
x_2 = \theta - \phi,
\end{equation}
\begin{equation}
y_2 = -(\theta + \phi).
\end{equation}
The integral is thus over $\Re^2\times T^2$.
Concentrate on the first 2 terms of the integrand involving derivatives with respect to  $x_1$ and $y_1$, the integrand is
\begin{equation}
\frac{1}{8}\int d\theta_zd\phi_wdy_1dx_1 \{\partial_{x_1}
     [(\sqrt{-g}g^{x\bar{x}})\partial_{x_1}\ln H] + \partial_{y_1}
     [(\sqrt{-g}g^{y\bar{y}})\partial_{y_1}\ln H] \}.
\end{equation}
The first term is zero since $H$ is independent of $x$ and we are left with the second term. Being a total derivative it reduces to
\begin{equation}\label{energy3}
\int dx_1d\theta_z d\phi_w [\sqrt{-g} g^{y\bar{y}} \partial_{y_1}\ln H]^\Lambda_{-\Lambda},
\end{equation}
where $\Lambda \to \infty$. 
In this limit where $y_1 \to \infty$,
\begin{equation}
(\sqrt{-g}g^{y\bar{y}})\partial_{y_1}\ln H \to 2\alpha(\frac{N}{12} -1).
\end{equation}

In the other limit where $y_1 \to -\infty$,
\begin{equation}
(\sqrt{-g}g^{y\bar{y}})\partial_{y_1}\ln H \to \alpha \partial_{y_1}(-2y_1 + \ln \frac{y_1}{2\pi}) \to -2\alpha.
\end{equation}
As a result (\ref{energy3}) becomes
\begin{equation}
\int dx_1(4\pi^2) \alpha (\frac{N}{6} -2 + 2).
\end{equation}
This diverges linearly in $x_1$ as expected due to the extra dimension
growing out of the intersection region. The extra factor of $4\pi^2$
originates from the two $SO(2)$ symmetries. We should rescale $x_1$ by
$\sqrt{\alpha}$ so that it is identified with the extra isometry direction.
A factor of $\sqrt{\alpha}$ is left in the integral. Thus $\alpha$ 
parametrises a family of different one-quarter BPS geometries. The
fact that the energy of a BPS solution depends on one or more
parameters (moduli) is familiar, for example, for the monopole in
Yang-Mills-Higgs theory.

\subsection{Some final remarks}
As we shall see in the next section, $j(\tau)$ can be equated to more general polynomials in $z$ and $w$.
\begin{equation}
j(\tau) = \frac{P(z,w)}{Q(z,w)}
\end{equation}As long as the polynomials are functions of $z^{N_1}w^{N_2}$ or one of the above generalised forms, we can solve them in exactly the same manner, except that the form of $F$ should be modified accordingly to
\begin{equation}
F = \eta(\tau)^2Q(z,w)^{-\frac{1}{12}}.
\end{equation}

As mentioned earlier the solution has not been generalised to the case
where both stacks of branes are separated. This is because if we alter
our ansatz to accommodate this situation by eq.(\ref{sep}) there are
no obvious new variables that would allow separation of variables in the equations of motion.

Solutions of intersecting D7 and $(p,q)$ 7-branes have been obtained
in \cite{Asano}, in which the complex modulus $\tau$ is arranged to
have dependence only on one of the complex coordinates, $z$. Using the
relation of compactified F-theory with type IIB on
orientifolds\cite{Sen1, Sen2}, the solution 
describes two orthogonal stacks of D7-branes and orientifold planes.
One stack consisting of sixteen D7-branes coinciding with four
orientifold planes is transverse to the $w$-plane. The other stack
consisting of another sixteen D7-branes and orientifold planes
separated,is transverse to the $z$-plane.
The form of these solutions is very similar to those explored here
except they are block 
diagonal in $z,w$. Yet our solutions have completely different
interpretations.  We 
have different monodromies that signify intersecting D7-branes separated
from any orientifold
 planes --- while not apparent in the $x,y$ coordinates  this could be
 read off from the $z,w$ coordinates. 
The fact that we could distinguish these solutions is encoded in the
different boundary
 conditions i.e. the compact directions in $x,y$ satisfy different
 periodicities.  

\section{Intersecting D7-branes and conifolds}\label{ftheory2}
\subsection{Conifolds and intersecting branes}
It was first observed in \cite{Vafa3} that a set of intersecting NS5
branes over 3+1 dimensions in type IIB is the T-dual of type IIA
compactified on a Calabi-Yau space. When the branes are arranged to 
intersect orthogonally and with no extra fields switched on, the
T-dual (along one of the totally transverse directions) geometry is a conifold at the intersection. The singularities
that occur in the dual geometry provide information of the gauge
symmetry the intersecting branes possess. There is an ADE
classification of singularities that could be described by 
algebraic varieties (see for example \cite{Vafa4}) which can be matched
with the Dynkin diagrams of groups. Therefore we can read off the kind
of enhanced gauge symmetry in the intersecting brane theory from the
geometric singularities that are present. The physical interpretation
of the symmetry enhancement is that the presence of a singularity
implies the collapse of some cycles upon which branes could wrap. As a
result the branes become massless and we obtain extra massless states
supplying the extra vector multiplets required for the symmetry.    

The intersecting D7-branes are intimately related to intersecting NS5
branes. Recall that the D7-brane solution is an adaptation of the
stringy cosmic string solution. Now suppose we go back to the original
context where the stringy cosmic string solution was considered,
i.e. type IIB compactified on a $T^2 \times M$ for some four
dimensional compact manifold $M$ to four dimensions. Starting with the
cosmic string solution with only a non-trivial complex modulus $\tau$
but trivial Kahler modulus $\rho = B + i\sqrt{G_{T^2}}$, where $B$ is
the NS 2-form and $G_{T^2}$ is the determinant of the metric on the
torus, we can perform a T-duality along one of the cycles on the
torus. This would exchange $\rho$ and $\tau$ and the monodromies in
$\tau$ would be transferred to $\rho$. As a result instead of an axion
magnetic charge we obtained magnetic charge of the NS 2-form potential
$B$. The magnetic charge is nothing other than the NS 5 brane in type
IIA (or we could equally well have started with cosmic strings in type
IIA). In the case of $N$ parallel cosmic strings the T-dual 5 brane
picture corresponds to $N$ parallel 5 branes transverse to the complex
$z$ plane and the torus $T^2$. The manner in which T- duality relates
geometry of type IIB (or IIA) with the NS5-brane configuration in type
IIA (or IIB) was studied in \cite{Vafa3}. The creation of
singularities by varying the complex parameters of the geometry that
leads to symmetry enhancement can now be understood from the T-dual
point of view as NS5 branes coinciding. 

It was then shown in \cite{Vafa3} that type IIB(A) on $T^2 \times T^2$
gives an analogous stringy cosmic string solution except that now
there are two tori moduli $\tau$ and $\tau'$ fibred on the complex $z$
plane and each can be described by a separate Weierstrass equation. If
the two tori become singular at the same point in the $z$ plane (say
$z=0)$, close to the singularity these equations are reduced to the
form 
\begin{eqnarray}
\zeta_1\zeta_2 &=& z \nonumber \\
\eta_1\eta_2 &=& z,
\end{eqnarray}
where $\zeta_i, \eta_j$ are all complex coordinates that parametrise
the two tori respectively. Combining the two equations we obtain 
\begin{equation}
\zeta_1\zeta_2 - \eta_1\eta_2 = 0
\end{equation}
which is the conifold equation. The corresponding NS5 brane
configuration is obtained by T-duality along a cycle of each torus as before, but
this time results in intersecting NS5 branes over 3+1
dimensions. Each 5 brane is transverse to one of the tori. That
intersecting singularities should lead to conifolds is quite a general
phenomenon. In this particular case it is the collision of two $A_0$
nodes that produces the conifold\footnote{It should be noted however
  that despite several attempts (e.g.\cite{coni1,coni2}), so far the conifold metric has not been satisfyingly obtained
  from intersecting 5-brane supergravity solutions.}.

Since intersecting D7-branes are basically the cosmic strings considered above the resultant geometry should carry intersecting singularities. Therefore it is very likely that we should be able to see a conifold from the F-theory picture, where there are enough dimensions to visualise the singularity.

\subsection{Intersecting D7-branes and F-theory}
The conifold is a Calabi-Yau three-fold. In order to see the conifold
that is allegedly lying at the intersection of two D7-branes, we need
two extra dimensions in addition to the four relatively transverse
dimensions. Therefore a discussion of the conifold only makes sense
when we can establish the precise relation between the ten-dimensional
type IIB with F-theory compactified on a Calabi-Yau.  The
compactification of F-theory on elliptically Calabi-Yau three-folds
was considered in \cite{Vafa2}. There are many different classes of
such Calabi-Yau three-folds. In the simplest case the base space is
the minimal ruled surface $\mathbb{F}_n$ i.e.  all possibilities where
a 2-sphere is fibred on another 2-sphere. The total space can again be
described by a Weierstrass equation as in (\ref{torus}) except that
now $f$ and $g$ are functions of two complex variables $z,w$. In
analogy to the case with parallel 7 branes  $f$ is a polynomial of
degree 8 w.r.t to $z$ and $w$ separately, and similarly for $g$, which is now a degree 12 polynomial in $z$ and $w$. In general they can be written as
\begin{eqnarray}\label{Weierstrass}
\alpha^2 &=& \beta^3 + f(z,w)\beta + g(z,w) \nonumber \\
f(z,w) &=& \sum_{i,j}^{8}  a_{ij}z^iw^j     \nonumber \\  
g(z,w) &=& \sum_{i,j}^{12}  b_{ij}z^iw^j
\end{eqnarray} 

This suggests that the type IIB interpretation should contain 24
7-branes intersecting another set of 24  7-branes orthogonally over
$5+1$ dimensions, which should give a compact space of $S^2$ fibred
over $S^2$. A related observation is that in the supergravity solution
we presented for $N_1=N_2=12$, the space becomes a product of two
cylinders. Therefore, as in \cite{cosmicstring}, when there are 24 branes on each stack we would glue the cylinders together to get two $S^2$, consistent with the F-theory picture.  
The zero locus of the discriminant of the Weierstrass should give the position of the branes. In general the branes wrap around curves in the $z$ and $w$ plane. 
The total number of complex moduli  modulo
SL($2,\mathbb{Z}$)$\times$SL($2,\mathbb{Z}$) and rescaling of $f$ and
$g$, is $9\times 9 + 13\times 13 -3-3-1 = 243$. This
corresponds to families of (3,243) Calabi-Yau manifolds. In general,
such F-theories should correspond to orientifold type IIB
theories\cite{Sen4}. We shall briefly review the arguments. To
see the orientifold planes and D7-branes, we first scale into a
particular region where $f$ and $g$ can be treated as polynomials of
degree (2,2) and (3,3), respectively, and where they
take special forms
\begin{eqnarray}
f(z,w) &=& C\eta(z,w)-2h(z,w), \nonumber \\
g(z,w) &=& h(z,w)(C\eta(z,w)-2h(z,w))
\end{eqnarray}
for some polynomials $\eta$ and $h$, and $C$ is a constant that is
tuned to be small such that from
\begin{equation}\label{jagain}
j(\tau) = \frac{4(24)^3(C\eta - 3h^2)^3}{C^2\eta^2(4C\eta - 9h^2)}
\end{equation} 
we can arrange $\tau \sim i\infty$ (i.e. the weak coupling limit in
which type IIB description makes sense).
Then the positions of D7-branes correspond to zeroes of the determinant $\Delta = C^2\eta^2(4C\eta - 9h^2)$, which gives
some colliding $A_1$ singularities (depending on the precise
form of $\eta$. e.g. for $\eta = z^2w^2$ we would have two colliding
$A_1$ singularities). Further, examining the monodromies of $\tau$
around $\eta$ gives an SL(2,$\mathbb{Z}$) transformation of
\begin{equation}
T^2 = \left(\begin{array}{cc}
1&1\\
0&1 \end{array}\right)^2 ,
\end{equation}
and so is consistent with the interpretation as a pair of coincident
(and perhaps curved in $z,w$ planes) D7-branes. The other zeroes of
$\Delta$ occur at
\begin{equation}\label{orientifoldplanes}
h(z,w) = \pm \frac{2}{3}\sqrt{C\eta(z,w)}.
\end{equation}
Again in the limit of small $C$ the zero locus is approximately $h =
0$ and we can look into the monodromies of $\tau$ when we move about
$h=0$. This would give the composition of the monodromies from each of
the two curves obtained in (\ref{orientifoldplanes}) and using (\ref{jagain}) gives an SL(2,$\mathbb{Z}$)
transformation, $\pm T^{-4}$. Therefore it is consistent to interpret
these curves (\ref{orientifoldplanes}) as two coincident orientifold
planes in weakly coupled type IIB. While we started off with some
products of SU(2) gauge symmetries (or $A_1$ singularities) we can
break these symmetries to obtain more general brane
configurations. This can be done by allowing $g$ to be
deformed to
\begin{equation}
g = h(C\eta - 2h^2) + C^2\chi,
\end{equation}
for some arbitrary polynomial $\chi$ subjected to the same restriction
as $g$. Again taking $C \to 0$ and $h,\eta,\chi$ fixed we go to the
weak coupling limit, and we
find that the discriminant becomes 
\begin{equation}
\Delta = -9C^2h^2(\eta^2 + 12h\chi).
\end{equation}
It is clear that the original coincident D7-branes given by $\eta = 0$
are now \emph{separated}. In this manner the breaking of the SU(2)
symmetry by deforming $g$ to obtain the most arbitrary configurations
is equivalent, from the type IIB picture, to the Higgs mechanism 
where branes move apart.
In order to reconstruct the complete $f$ and $g$ that describe all
the branes we would need to patch different regions which locally
admit such a description. One example is the correspondence between
F-theory on $\mathbb{F_0}$ (where the base space is simply the direct product
of two 2-sphere) and the T-dual of the Gimon-Polchinski
model\cite{Sen2,GP,Sen3}. In this correspondence it was observed that
the number of constraints needed to obtain an
(SU$(2)$$)^8\times$(SU$(2)$$)^8$ gauge symmetry exceeds the total
number of complex moduli. Fortunately there are still families of
solutions that could satisfy all the constraints but these are constrained to lie in a particular slice of the moduli space. As a
result the total number of massless hypermultiplets is less than 243. But
the number of extra neutral hypermultiplets that appear as  SU(2) is broken matches the F-theory prediction as the constraints on $f$ and $g$ are lifted. 
Returning to the question of the search for the conifold, that arises
from the orthogonal intersection of two D7-branes, we could analyse
the geometry by scaling close to the intersection region while
arranging the orientifold planes to be sufficiently far so that they
do not interfere with the local geometry. This will be carried out in the next section.

\subsection{Conifold in F-theory}
Having now reviewed the relation between F-theory compactified on an
elliptically fibred Calabi-Yau three-fold with type IIB orientifold
theory, the various parameters of the Weierstrass equation can now be
varied to build a conifold in F-theory which can be interpreted as
orthogonally intersecting D7-branes in type IIB.  
For this to happen two $A_0$ nodes corresponding to $U(1)\times U(1)$ gauge symmetry must collide.
From (\ref{torus}) it is straight forward to find the specific forms
of $f$ and $g$ needed to construct these colliding $A_0$
singularities,  
\begin{eqnarray}
f &=& a_{00} + a_{11}zw + O(z^2,w^2) \nonumber \\
g &=& b_{00} + b_{11}zw + O(z^2,w^2). 
\end{eqnarray}
The discriminant is then
\begin{equation}\label{Delta}
\Delta = 4f^3+27g^2 = (4a_{00}^3+27b_{00}^2) + zw[A + O((zw)^2)],
\end{equation}
where $A$ is just some combination of $a_{11}$ and $b_{11}$.
In order to describe colliding $A_0$ singularities at $z=w=0$ we need
to set
\begin{equation}
(4a_{00}^3+27b_{00}^2)=0,
\end{equation}  and
\begin{eqnarray}
\alpha &=& 0 \nonumber \\
\beta &=& \sqrt{-\frac{a+zw+O(z^2w^2)}{3}}.
\end{eqnarray}
 These are the points where (\ref{torus}) is satisfied  and all its partial
derivatives vanish. They are the singular points. Expanding
(\ref{torus}) about these singular points up to second order, we get
\begin{equation}
\delta_\beta^2 - \delta_\alpha^2 + b_{00}zw = 0,
\end{equation}
which is the conifold equation. To summarise, the conifold equation is built from two
requirements that are necessary if it is to represent intersecting
D7-branes. First, the zero locus of $\Delta$, which is interpreted as
the position of the D7's, should represent two
orthogonally intersecting $A_0$ nodes. Second, the monodromies about
these 7-branes should correspond to those of D7-branes. 

The next obvious question to ask is
then whether we could obtain the conifold metric from our metric for intersecting D7-branes by applying the
standard procedure described in \cite{cosmicstring}, which extends the
lower dimensional metric to a higher dimensional one including the
torus. This presents difficulties, as we will now see.

\subsection{Obstruction to obtaining the conifold metric from intersecting 7 branes solution}
We would like to extend our solution of the ten dimensional Einstein
equations to twelve dimensions in order to visualise the
conifold\footnote{Certainly we should be very cautious with this
  search for metrics because the compact torus in F-theory is actually
  1+1 dimensions and yet we are treating them as purely
  Euclidean.}. It was argued in \cite{cosmicstring} that if the total
space $E$, which is the elliptically fibred Calabi-Yau $n$-fold
F-theory is compactified on, admits a nowhere vanishing holomorphic
$(n,0)$ form $W$ and that we can put a Kahler metric on $E$ whose
corresponding volume form approaches $W\wedge \bar{W}$ asymptotically,
then $E$ admits a Ricci-flat Kahler metric for non-compact
$E$\cite{cosmicstring}. Since we are scaling into the conifold it is
reasonable to consider $E$ to be non-compact. Following these lines
the corresponding $(3,0)$ form $W$ in our solution would be 
\begin{equation}
W = \Delta^{-\frac{1}{12}}\eta^2(\tau)d\zeta dx dy,
\end{equation}
where $\zeta$ is a complex coordinate on the torus with periodicity
\begin{equation}
\zeta \sim \zeta + 1 \sim \zeta + \tau,
\end{equation}
and $\Delta \sim (zw[A + O((zw)^2)$ is as defined in eq. (\ref{Delta}).

The Kahler potential for this extra torus which has all the required
properties, including invariance under $SL(2,\mathbb{Z})$
transformation, is then given by 
\begin{equation}
\Phi_{T^2} = -\frac{(\zeta - \bar{\zeta})^2}{2\tau_2}.
\end{equation}
This part of the potential gives the metric on the F-theory torus,
\begin{equation}\label{torusmetric}
ds_{T^2}^2 = \frac{1}{\tau_2}|d\zeta - \frac{\zeta - \bar{\zeta}}{2i\tau_2}\partial_y \tau dy|^2.
\end{equation}
This metric is automatically Kahler flat since the 12-dimensional Ricci scalar would be
\begin{equation}
R = G^{a\bar{b}}\partial_a\partial_{\bar{b}} (\ln g_{10} - \ln \tau_2) = 0
\end{equation}
where $g_{10}$ is the determinant of the 10-dimensional Kahler metric and we immediately identify what appears in the bracket as the 10-dimensional Einstein equation. 
Now take the limit $zw \to 0$ or $y_1 \to \infty$, then
\begin{eqnarray}
\tau &\sim& \frac{y}{2\pi i} \nonumber \\
K_{y\bar{y}} &\sim& \exp{(-2y_1)}\frac{y_1}{8\pi}
\end{eqnarray}

The total Kahler potential  in this limit is (keeping only leading order terms)
\begin{equation}\label{kahlerpot}
K \sim \frac{2\zeta_2^2}{y_1} + |x|^2 -  \frac{e^{-2y_1}y_1}{8\pi}.
\end{equation}
Note that the third term is exponentially suppressed compared to the first two terms.
We wish to compare this Kahler potential with that of the conifold. The Kahler potential of a conifold is\cite{candelas}
\begin{equation}
\Phi_{\mathrm{conifold}} = \frac{3}{2}r^{\frac{4}{3}}
\end{equation}
where 
\begin{eqnarray}
a^2 + b^2 + c^2 + d^2 &=& 0 \nonumber \\
|a|^2 + |b|^2 + |c|^2 + |d|^2 &=& r^2.
\end{eqnarray}
In order to compare the two Kahler potentials it would be more convenient to change coordinates to $\zeta, x,y$. 
The torus coordinate $\zeta$ is related to the coordinates in the Weierstrass equation (up to scaling) (\ref{Weierstrass}) by
\begin{equation}
\zeta = \int_{\beta_0}^{x} \frac{d\beta}{\alpha},
\end{equation}
along some path $C$ in the complex $\beta$ plane. Since we are now
restricting to a region very close to the conical singularity $
\delta_\alpha^2= \delta_\beta^2 =zw = 0 $ we can ignore $\delta_\beta^3$ and so we obtain
\begin{equation}
\zeta = \ln (\delta_\alpha + \delta_\beta) - \frac{1}{2}\ln zw.
\end{equation}
Substituting into eq. (\ref{Weierstrass}) we have
\begin{eqnarray}\label{r}
\delta_\alpha &=& \sqrt{zw} \cosh \zeta, \nonumber \\
\delta_\beta &=& \sqrt{zw} \sinh \zeta,  \nonumber \\
r^2 &=& |zw|\cos 2\zeta_2 + \frac{|z|^2 + |w|^2}{2} = e^{-y_1}[\cosh x_1 + \cos 2\zeta_2].
\end{eqnarray}
Now the Kahler potential as a function of $r$ can be expressed in our coordinates and the resulting metric can be shown to be Ricci flat, which is very much in the same spirit as \cite{conifold1}. We first solve the conifold equation to obtain the local coordinates on the conifold and then impose an SO($n$) symmetry by obtaining the expression of $r$ such that the Kahler potential is solely a function of $r$. Finally we can compute the Ricci scalar and set it to zero. The Kahler potential is then obtained from this differential equation.
Comparing (\ref{r}) with the D7-brane Kahler potential
(\ref{kahlerpot}) we are impressed that both depend on the same
coordinates $y_1, \zeta_2, x_1$ but (\ref{kahlerpot}) depends also on
$x_2$. Considering the discussion in section {\ref{generalsols}} a possible modification to our Kahler potential could be
\begin{equation}
K_{\mathrm{modified}} = \frac{2\zeta_2^2}{y_1} + \frac{x_1^2}{y_1} -  \frac{e^{-2y_1}y_1}{8\pi}.
\end{equation}
Ignoring also the third term since it is exponentially suppressed, we have
\begin{equation}
K_{\mathrm{modified}} \sim  \frac{2\zeta_2^2}{y_1} + \frac{x_1^2}{y_1},
\end{equation}
which now has a more similar structure to (\ref{r}) and indeed is
always positive and approaches zero as $zw \to 0$, and can therefore
be identified with some radius of the conifold. However, it is
intrinsically different from the conifold Kahler potential by its
dependence on $y_1$, which cannot be remedied by a different choice of
metric, in the base space. This is rooted in the form of the metric,
which is block diagonal in the torus metric and the base space,
breaking explicitly the required SO(4) symmetry of the conifold. This
can be seen by a coordinate transformation 
\begin{equation}
\zeta = t_1 + \tau t_2
\end{equation}
where $t_i \sim t_i + \mathbb{Z}$ and $t_i \in \mathbb{R}$. The torus metric (\ref{torusmetric}) then becomes
\begin{equation}
ds_{T^2}^2 = \frac{1}{\tau_2} |dt_1 + \tau dt_2|^2,
\end{equation}
which is the standard torus metric with complex modulus $\tau$ and unit metric determinant, consistent with the fact that the torus is unphysical since its Kahler modulus is not part of the physical spectrum and cannot be varied.
The same problem also plagues the metric of K3 with $A_1$ singularity
in the parallel stringy cosmic string solution. In that case the SO(3)
symmetry expected of a the conical singularity is again broken by the
choice of the block diagonal ansatz.  

This is similar to the observation in \cite{vafabh} where it is found
that the T-dual to the stringy cosmic string solution does not give
the exact NS5 brane metric because of the block diagonal ansatz of the
metric in the torus and the transverse complex plane $z$. It was
argued there that  the correct solution should emerge if we go beyond
the adiabatic approximation used in \cite{cosmicstring}. However,
given that the extra dimensions are not physical and no momentum can
propagate along those directions, it is questionable as to whether it
is sensible to  obtain a metric which preserves an SO(4) symmetry
mixing these directions with those along the base space.  
Another observation is that in \cite{vafainstanton} the leading order
metric of the moduli space of type IIA on a Calabi Yau three-fold near
the conifold point is given by the stringy cosmic string solution.  On
the other hand we have seen here how the stringy cosmic string 
possibly describes a conifold directly, albeit with the need for
some modifications. This seems to suggest some curious connection
between a geometry and its moduli space.  

Finally we would like to make a remark about deformed and resolved
conifolds. From the derivation of (\ref{Weierstrass}) the deformed
conifold equation 
\begin{equation}
a^2 + b^2 + c^2 + d^2 = \epsilon
\end{equation}
for small $\epsilon$ can simply be obtained by replacing $zw$ by $zw -
\epsilon$ everywhere. This would correspond to curved  7-branes, very
much analogous to the NS5 brane configurations dual to a deformed
conifold. However, learning from the NS5 story, we know that the
deformed conifold is related to a resolved conifold by mirror symmetry
via three T-dualities along three isometry directions \cite{5branes},
or by a conifold transition. So do our curved D7-branes producing the
deformed conifold T-dualise to a resolved conifold? Also is there some
D7-brane configuration which naturally produces a resolved conifold?
The answer is probably no to both questions. This is because, as is
well known, a resolved conifold is obtained from the conifold by
varying the Kahler moduli, which are not physical modes in
F-theory. The Kahler modulus of the torus is a priori fixed. From
another perspective, in order to go over to the mirror manifold we
have to perform three T-dualities and this must involve T-duality
along one of the cycles of the fibred torus. Since no physical modes
can propagate in the torus, it is not know whether T-duality along
these directions makes any sense.

\section{Discussion and Conclusion}
In this paper we have obtained  several supergravity solutions that 
describe intersecting  D7-branes. The solutions turn out to be very
similar to parallel D7-brane solutions and the major difference
lies in the non-trivial boundary conditions. This suggests that an
enhancement of supersymmetry away from the singular point of the
geometry where the branes reside is a quite generic phenomenon and
it seems that the symmetry enhancement observed in intersecting
D5-branes in \cite{IKS} also occurs for intersecting D7's.  

However, there is an important difference between intersecting D5's in
\cite{IKS} and D7 since the connection between the microscopic string
description and supergravity is unclear for a system of
D7-branes. It is impossible to take the large-$N$ limit since each
D7-brane induces a deficit angle of $\pi/6$ independent of the string
coupling. It is therefore impossible to ignore the back-reaction of
the branes on space-time. This is seen in the string description from
the fact that closed string exchange cannot be ignored. At the very
least, it is necessary to include the coupling of the RR-scalar,
$\tau_1$, to the bulk Riemann tensor\cite{green}. As a result 
a probe brane analysis using the DBI action could be quite
subtle\cite{tseytlin}. 

On the other hand, the geometry generated by intersecting D7-branes
is of interest in its own right. For example the world-volume theory of a D3
brane embedded in the intersection domain of two stacks of D7-branes
is $N=1$ QCD with flavors \cite{holography}. The supergravity solution of
intersecting D7-branes discussed in this work should be a good
starting point for exploring the gravity dual of $N=1$ QCD.

By exploring further the relations of 7-branes in type IIB with
F-theory, it looks very likely that a conifold should emerge at the
intersection of two orthogonal D7-branes from the 12-dimensional
F-theory picture. However, we haven't been able to obtain the explicit
conifold metric by extending our solution to twelve dimensions
applying the procedure described in \cite{cosmicstring}. This is due
to the breaking of the SO(4) symmetry of the conifold by the block
diagonal ansatz of the metric which distinguishes the fibred torus from
the base space. This breaking of symmetry is possibly intrinsic to
F-theory since the extra dimensions are unphysical and it is
hard to imagine how they could have mixed with the other physical
directions\footnote{In fact the search for such a metric might not make
sense in the first place because these extra dimensions are not even
Euclidean.} The deformed conifold can be obtained by turning the D7
branes into curved ones, very much analogous to the NS5 brane
story. However, it is unlikely that there is a resolved conifold
counterpart in the D7 picture. This is once again due to the
unphysical nature of the extra dimensions, whose only remnant after
BRST projection is the complex modulus $\tau$. The Kahler modulus is
non-existent and it would not be possible to resolve the singularity
by modifying the Kahler modulus. From the brane perspective, again
comparing with the NS5 branes, D7-branes separated along the totally
transverse directions would probably produce a resolved conifold. Yet
these totally transverse directions are exactly those non-physical
torus directions. The D7's are thus stuck together rendering a
resolved conifold improbable.

Finally there are several clear aspects of these kinds of intersections that
have not been explored.
We would like to understand more about the proposed solution in the
case where $N_1 \ne N_2$ and the other families of solutions giving
non-block-diagonal solutions.

We should understand how angle deficits
can be evaluated in these situations and the way to
generalise to non-coincident branes in both stacks or possibly other
curved brane configurations. 
It would be interesting to see how the symmetry enhancement seen in
the supergravity solution can be described in microscopic terms by
string theory. 
It would also be interesting to understand if the conifold plays
any role in these intersecting D7's and if there exists more general
null projection in F-theory that could reproduce the 10 dimensional
IIB metric from a twelve-dimensional starting point.

\section*{Acknowledgement}
I would like to express my gratitude for the support, guidance and
encouragement given to me by Prof. M. Green. I would also like to
thank Prof. D. Kutasov for pointing out the analogy between
intersecting D7-branes and NS5-branes and the connection with the conifold.

\appendix

\section{An overview of stringy cosmic strings, D7-branes and their
  relation to F-theory}

\subsection{Stringy cosmic string and D7-branes}\label{cosmicsolun}

The supergravity solution of a cosmic string in four dimensions which induces non-trivial SL($2,\mathbb{Z}$) monodromies on
the complex modulus $\tau $ was considered in \cite{cosmicstring}. The
complex modulus $\tau$ originates from a torus on which a higher
dimensional theory compactifies, therefore the theory should be
invariant under SL($2,\mathbb{Z}$) transformation, where 
\begin{equation}\label{SL2z}
\tau \to \frac{a\tau + b}{c\tau + d}
\end{equation}
and $ad- bc = 1$, $a,b,c,d$ are integers.  In type IIB string theory
it is believed that SL($2,\mathbb{Z}$) is an exact symmetry even in
the quantum theory.  The complexified string coupling combining the RR
zero form and the string coupling, also usually known as $\tau = C_0 +
ie^{-\phi}$ transforms as in (\ref{SL2z}) and the form of the IIB
effective action looks exactly like the compactified supergravity
theory mentioned above\cite{cosmicstring}. This suggests that the
solution can be adopted for describing other extended objects of
codimension 2, which are  D7-branes in ten dimensions \cite{GGP}.   

The bosonic part of type IIB supergravity is given by

\begin{equation}
\int \sqrt{g}(R + \frac{\partial_\mu \tau\partial_\nu\bar{\tau}}{2\tau^2_2}g^{\mu\nu})
\end{equation}
where only the dilaton and $RR$ zero form are switched on.

The equations of motion obtained are: \newline
1. The $\tau$ equation:
\begin{equation}\label{tau}
\partial_\mu(\sqrt{-g}g^{\mu\nu}\partial_\nu\tau)= 2\sqrt{-g}g^{\mu\nu}\frac{\partial_\mu\tau\partial_\nu\tau}{\tau-\bar{\tau}}.
\end{equation} 
\newline
2. Einstein equation:
\begin{equation}\label{einstein}
R_{\mu\nu}=\frac{1}{4\tau^2_2}(\partial_\mu\tau\partial_\nu\bar{\tau} + \partial_\nu\tau\partial_\mu\bar{\tau} ).
\end{equation}
Note also that there is a composite $U(1)$ connection 
\begin{equation}
Q_\mu = i\frac{\partial_\mu \tau_1}{\tau_2}
\end{equation}
that enters into covariant derivatives along with the spin connection.
Suppose the D7-branes are aligned to be transverse to $x^8$ and $x^9$. It will be more convenient to combine these two transverse dimensions into a complex coordinate $z = x^8 + ix^9$ since the resultant geometry is Kahler. All fields depend only on $z$.  Substituting the metric ansatz
\begin{equation}
ds^2 = \eta_{\mu\nu}dx^\mu dx^\nu + \Omega^2(z)dzd\bar{z}.
\end{equation}
into (\ref{einstein}) gives
\begin{equation}
2\partial\bar{\partial}\ln \Omega = \partial\bar{\partial}\ln \tau_2.
\end{equation}
It turns out that any holomorphic $\tau$ satisfies the equations and a convenient solution for the metric is
$\Omega = \tau_2$.  However this is not SL($2,\mathbb{Z}$) invariant and an acceptable solution is
\begin{eqnarray}
j(\tau) &=& \frac{1}{z}  \label{jeqn}\\
\Omega^2 &=& \tau_2|\eta^2(\tau)z^{-1/12}|^2, 
\end{eqnarray} 
where $j(\tau)$ maps the fundamental region in the $\tau$ plane to a complex plane exactly once. In fact
\begin{equation}
j(\tau)= \frac{\theta_2(\tau)^8+\theta_3(\tau)^8+ \theta_4(\tau)^8}{\eta^{24}}.
\end{equation} 
The dedekind eta function is a modular function that compensates for the transformation in $\tau_2$. The factor of $z^{-1/12}$ removes the singularity in the eta function as $z$ approaches zero.
It follows from (\ref{jeqn}) that $\tau \to \infty$ as $z \to 0$ and that $\tau \to \tau + 1$ as $z\to ze^{2\pi i}$, signifying the presence of a D7-brane at $z=0$.
As $z \to \infty$, $\tau$ approaches a constant value and the metric becomes
\begin{equation}
\Omega^2 = |z^{-1/12}dz|^2,
\end{equation}
which implies an angle deficit of $\pi/6$. The solution generalises to describe more branes, by equating $j(\tau)$ to more general holomorphic functions in $z$. Since each brane produces an angle deficit of $\pi/6$, 24 parallel branes have angle deficit of $4\pi$, in which case the transverse space is compactified to a sphere.  
In general the presence of angle deficits breaks all
supersymmetries, as can be seen from the killing spinor equation, but the D7-brane solution is supersymmetric. This property follows from a cancellation of the effect of the spin connection by the $U(1)$ connection. 

\subsection{Relation to F-theory}
As discussed above, in the presence of 24 branes
the transverse space is compactified to a two sphere. In the stringy
cosmic string context there is a torus whose modulus is $\tau$ fibred
over this sphere, where the total space is a K3 surface. However in
the description of 7-branes $\tau$ is the complexified string
coupling and there are not any torus related to $\tau$. The analogy is
so strong that it was suggested in \cite{Vafa1} that type IIB with 24
7-branes is equivalent to
twelve dimensional F-theory compactified on an elliptically fibred
$K3$\cite{Vafa1}. The modulus of the fibre, the torus, is $\tau$. The
two extra dimensions, however, are not dynamical and by a suitable
projection operator no physical states propagate along these extra
dimensions and the only remnant is $\tau$ \footnote{There is another
  complication about the signature of these extra dimensions being
  (1,1). But as explained in \cite{Vafa1} we can treat it as a
  Euclidean space.}. An elliptically fibred $K3$ can be conveniently described by a Weierstrass equation whose coefficients are functions of the base space, i.e.
\begin{equation}\label{torus}
y^2 = x^3 + f(z)x + g(z).
\end{equation}
The zero locus of the discriminant of the equation
\begin{equation}\label{delta}
\Delta = 4f^3 + 27g^2 
\end{equation}
gives the position of the 7 branes and the modulus of the torus is related to $f$ and $g$ by
\begin{equation}
j(\tau) = \frac{(24f)^3}{\Delta}.
\end{equation}
For equation (\ref{delta}) to describe 24 branes it  should have in general
twenty-four distinct solutions. If $f$ is a polynomial in $z$ of
degree 8 and $g$ of degree 12, modding out by SL($2,\mathbb{C}$)
symmetry on the $z$ plane and the freedom to rescale $f$ and $g$, the
number of independent complex parameters is 18. This tells us that there cannot be more than eighteen 7-branes of the same type. This is
inevitable since the theory on a compact manifold would have been inconsistent had
fluxes not cancelled.  It was shown in \cite{Sen1} that type IIB on
an orientifold $T^2/\mathbb{Z}_2$ with four orientifold 7-planes, each
coinciding with four D7-branes, is F-theory on K3 at a particular
point of the complex moduli space where the modulus $\tau$ is a
constant. The charges of the D7-branes and the O7-planes cancel
exactly everywhere and the number of massless gauge neutral fields is
exactly 18. They correspond to the modulus $\tau$, 16 massless fields
that give the D7-brane position and $4-3=1$ orientifold positions
modulo $SL(2,\mathbb{C}$) on the $z$ plane.  
Analogous considerations for intersecting branes are discussed in the main text.


\begin{thebibliography}{}
\bibitem{green}
 M.~B.~Green, J.~A.~Harvey and G.~W.~Moore,
``I-brane inflow and anomalous couplings on D-branes,''
  Class.\ Quant.\ Grav.\  {\bf 14} (1997) 47
  [arXiv:hep-th/9605033].

\bibitem{IKS} 
N.~Itzhaki, D.~Kutasov and N.~Seiberg,
  ``I-brane dynamics,''
  JHEP {\bf 0601} (2006) 119
  [arXiv:hep-th/0508025].

\bibitem{cosmicstring} 
B. Greene, A. Shapere, C. Vafa, S. Yau, 
"Stringy Cosmic Strings And Noncompact Calabi-Yau Manifolds," 
 Nucl.\ Phys.\ B \textbf{B337} (1990) 1


\bibitem{Vafa3}
 M.~Bershadsky, C.~Vafa and V.~Sadov,
  ``D-Strings on D-Manifolds,''
  Nucl.\ Phys.\ B {\bf 463} (1996) 398
  [arXiv:hep-th/9510225].


\bibitem{Brandeis}
 M.~Kruczenski,
  ``Supergravity backgrounds corresponding to D7 branes wrapped on Kaehler
  manifolds,''
  JHEP {\bf 0401} (2004) 031
  [arXiv:hep-th/0310225].



\bibitem{Asano}
  M.~Asano,
  ``Stringy cosmic strings and compactifications of F-theory,''
  Nucl.\ Phys.\ B {\bf 503} (1997) 177
  [arXiv:hep-th/9703070].


\bibitem{Sen1}
 A.~Sen,
  ``F-theory and Orientifolds,''
  Nucl.\ Phys.\ B {\bf 475} (1996) 562
  [arXiv:hep-th/9605150].

\bibitem{Sen2}
 A.~Sen,
  ``A non-perturbative description of the Gimon-Polchinski orientifold,''
  Nucl.\ Phys.\ B {\bf 489} (1997) 139
  [arXiv:hep-th/9611186].


\bibitem{Vafa4}
 M.~Bershadsky, K.~A.~Intriligator, S.~Kachru, D.~R.~Morrison, V.~Sadov and C.~Vafa,
  ``Geometric singularities and enhanced gauge symmetries,''
  Nucl.\ Phys.\ B {\bf 481} (1996) 215
  [arXiv:hep-th/9605200].


\bibitem{Vafa2}
D.~R.~Morrison and C.~Vafa,
  ``Compactifications of F-Theory on Calabi--Yau Threefolds -- I,''
  Nucl.\ Phys.\ B {\bf 473} (1996) 74
  [arXiv:hep-th/9602114],
  ``Compactifications of F-Theory on Calabi--Yau Threefolds -- II,''
  Nucl.\ Phys.\ B {\bf 476} (1996) 437
  [arXiv:hep-th/9603161].



\bibitem{coni1}
  K.~Dasgupta and S.~Mukhi,
  ``Brane constructions, conifolds and M-theory,''
  Nucl.\ Phys.\ B {\bf 551} (1999) 204
  [arXiv:hep-th/9811139].

\bibitem{coni2} 
 K.~Ohta and T.~Yokono,
  ``Deformation of conifold and intersecting branes,''
  JHEP {\bf 0002} (2000) 023
  [arXiv:hep-th/9912266].




\bibitem{Sen4} 
 A.~Sen,
  ``Orientifold limit of F-theory vacua,''
  Phys.\ Rev.\ D {\bf 55} (1997) 7345
  [arXiv:hep-th/9702165].



\bibitem{GP}
E.~G.~Gimon and J.~Polchinski,
  ``Consistency Conditions for Orientifolds and D-Manifolds,''
  Phys.\ Rev.\ D {\bf 54} (1996) 1667
  [arXiv:hep-th/9601038].




\bibitem{Sen3}
 A.~Sen,
  ``F-theory and the Gimon-Polchinski orientifold,''
  Nucl.\ Phys.\ B {\bf 498} (1997) 135
  [arXiv:hep-th/9702061].



\bibitem{candelas} 
 P.~Candelas and X.~C.~de la Ossa,
  ``COMMENTS ON CONIFOLDS,''
  Nucl.\ Phys.\ B {\bf 342} (1990) 246.


\bibitem{conifold1}
R.~Parthasarathy and K.~S.~Viswanathan,
  ``Non-linear sigma model on conifolds,''
  Mod.\ Phys.\ Lett.\ A {\bf 17} (2002) 517
  [arXiv:hep-th/0111097].


\bibitem{vafabh} 
 H.~Ooguri and C.~Vafa,
  ``Two-Dimensional Black Hole and Singularities of CY Manifolds,''
  Nucl.\ Phys.\ B {\bf 463} (1996) 55
  [arXiv:hep-th/9511164].



\bibitem{vafainstanton} 
H.~Ooguri and C.~Vafa,
  ``Summing up D-instantons,''
  Phys.\ Rev.\ Lett.\  {\bf 77} (1996) 3296
  [arXiv:hep-th/9608079].





\bibitem{5branes}
M.~Aganagic, A.~Karch, D.~Lust and A.~Miemiec,
  ``Mirror symmetries for brane configurations and branes at singularities,''
  Nucl.\ Phys.\ B {\bf 569} (2000) 277
  [arXiv:hep-th/9903093].



\bibitem{tseytlin}
  A.~A.~Tseytlin,
  ``*No-force* condition and BPS combinations of p-branes in 11 and 10
  dimensions,''
  Nucl.\ Phys.\  B {\bf 487} (1997) 141
  [arXiv:hep-th/9609212].


\bibitem{holography}
  B.~A.~Burrington, J.~T.~Liu, L.~A.~Pando Zayas and D.~Vaman,
  "Holographic duals of flavored N = 1 super Yang-Mills: Beyond the probe
  approximation,"
  JHEP {\bf 0502} (2005) 022
  [arXiv:hep-th/0406207].



\bibitem{GGP}
  G.~W.~Gibbons, M.~B.~Green and M.~J.~Perry,
  ``Instantons and Seven-Branes in Type IIB Superstring Theory,''
  Phys.\ Lett.\ B {\bf 370} (1996) 37
  [arXiv:hep-th/9511080].

\bibitem{Vafa1}
 C.~Vafa,
  ``Evidence for F-Theory,''
  Nucl.\ Phys.\ B {\bf 469} (1996) 403
  [arXiv:hep-th/9602022].



\end{thebibliography}
\end{document}